\documentclass[aip,jap,preprint]{revtex4-1}
\usepackage{graphicx,natbib,amssymb}
\usepackage{dcolumn}%
\usepackage{bm}%

\begin{document}
\newcommand{\cm}{cm$^{-1}$ }
\newcommand{\oC}{$^{\circ}$C }
\newcommand{\BZN}{Ba(Zn$_{1/3}$Nb$_{2/3}$)O$_3$ }
\newcommand{\complex}{A(B$'_{1/3}$B$''_{2/3}$)O$_3$ }
\newcommand{\tri}{$P\overline{3}m1$ }
\newcommand{\pero}{$Pm\overline{3}m$ }

\title{Ba(Zn$_{1/3}$Nb$_{2/3}$)O$_3$ thin films obtained by polymeric precursors method}
\date{October, 10, 2012}

\author{Jo\~ao Elias Figueiredo Soares Rodrigues}
\affiliation{Departamento de F\'{\i}sica, CCET, Universidade Federal do Maranh\~ao, 65085-580, S\~ao Lu\'{\i}s - MA, Brazil}
\author{D\'ebora Morais Bezerra}
\affiliation{Departamento de Qu\'{\i}mica, CCET, Universidade Federal do Maranh\~ao, 65085-580, S\~ao Lu\'{\i}s - MA, Brazil}
\author{Adeilton Pereira Maciel}
\affiliation{Departamento de Qu\'{\i}mica, CCET, Universidade Federal do Maranh\~ao, 65085-580, S\~ao Lu\'{\i}s - MA, Brazil}
\author{Alexandre Rocha Paschoal}
\affiliation{Departamento de F\'{i}sica, Universidade Federal do Cear\'a, Campus do Pici, 60455-760, Fortaleza - CE, Brazil}
\author{Carlos William de Araujo Paschoal}
\affiliation{Departamento de F\'{\i}sica, CCET, Universidade Federal do Maranh\~ao, 65085-580, S\~ao Lu\'{\i}s - MA, Brazil}
\email{paschoal@ufma.br}
\email{paschoal.william@gmail.com}
\thanks{Author whose correspondence should be addressed: Phone +55 98 3301.9209; Fax: +55 98 3301 8204}

\begin{abstract}
In this work \BZN thin films were prepared by the polymeric precursors method. A detailed description of the procedure to obtain the precursors and deposition was presented. High quality polycrystalline films of \BZN were obtained from Pt/Ti/SiO$_2$/Si substrate deposited by spin coating technique. X ray measurements show that the film crystallizes in a partial ordered trigonal structure characteristic of the 1:2 complex perovskites. The partial order and structure was confirmed by Raman measurements. Also, confocal Raman and atomic force microscopy was used to characterize the film morphology.
\end{abstract}


\maketitle

\section{Introduction}
Due to their high selectivity and bandwidth as well as reduced size for a given resonant frequency, dielectric ceramics have been fully employed to manufacture filters to be used in wireless telecommunication systems \cite{WersingW.1996,Sebastian2008}. Typical parameters for this application are a high unload quality factor $Q_u$ higher than 20,000 at 2 GHz, which imply in selectivity; a static permittivity $\varepsilon_o$ higher than 30, for miniaturization; and a temperature coefficient of the resonant frequency $\tau_F$ tunable through zero, for thermal stability \cite{WersingW.1996,Moulson2003}. Ba-based ceramics BaTi$_4$O$_9$ and Ba$_2$Ti$_9$O$_{20}$ were the first dielectric ceramics whose physical properties fulfil these technical requirements. \cite{Masse1971,PLOURDE1975} Today, the main compounds employed are based on CaTiO$_3$$-$NdAlO$_3$ and ZrTiO$_4$$-$Zn$_2$Nb$_2$O$_7$ solid solutions, which have, typically, $\varepsilon_o \approx 45$ and $Q_u \approx 25,000$ at 2 GHz. \cite{Seiichiro1994,Okuyama1994}.

Ta-based ceramics as Ba(Zn$_{1/3}$Ta$_{2/3}$)O$_3$ (BZT)\cite{nomura1983,Sagala1993a} and Ba(Mg$_{1/3}$Ta$_{2/3}$)O$_3$ (BMT) \cite{BarberD.J.MouldingK.M.Zhou1997} show very attractive $Q_u$ values higher than 70,000 at 2 GHz that can increase the selectivity and optimize the bandwidth. However, the high cost of the Ta$_2$O$_5$ start oxide used in preparation by solid state route of these Ta-based ceramics increases the filter costs a lot \cite{Pullar2009}. Since Nb and Ta ions has the same ionic radii and Nb$_2$O$_5$ is isostructural to and cheaper than Ta$_2$O$_5$, it is an excellent alternative way is to obtain dielectric ceramics based on the substitution of Ta by Nb. Thus, several scientific investigations on the microwave properties of Nb-based dielectric ceramics have been performed \cite{Dias2003b,Ahn2002,Akbas2005,BarberD.J.MouldingK.M.Zhou1997,Chaouchi2009,Chen2006,Dias2001,Dias2007,Kawashima1977,KimB.K.HamaguchiH.KimI.THong1995,KimI.T.Kim1997,Lee2007,Noh2002,Sert2009a,Solomon2011,Thirumal2002,Veres2005,Yin2004,Yue2004}.
So, despite some problems in achieving zero $\tau_F$ and optimizing $Q_u$, \BZN (BZN) has great potential as the starting point for the development of dielectric ceramics for low cost devices \cite{Hughes2001a}.

Complex perovskites A(B$'_{1/3}$B$''_{2/3}$)O$_3$, as Ba(Zn$_{1/3}$Nb$_{2/3}$)O$_3$, can be disordered or ordered according to the B$'$ and B$''$ ion site distributions. When B$'$ and B$''$ are randomly distributed into the B-site of the simple ABO$_3$ perovskite the compounds are disordered, crystallizing in a cubic structure with $Pm\overline{3}m$ symmetry. However, when B$'$ and B$''$ are alternately distributed to the same site in a sequence $\cdot\cdot\cdot$B$'$B$''$B$''$B$'$B$''$B$''$$\cdot\cdot\cdot$ (called 1:2 ordering), the structure crystallizes in a trigonal symmetry belonging to the $P\overline{3}m1$ space group. This order occurs along the $\left < 111 \right > $ direction and is very important to obtain high $Q_u$ and low $\tau_F$.\cite{Davies1999}. Despite several investigations in BZT and BZN compounds in bulk form \cite{ZhouJ.Barber1997}, few works described the BZT film synthesis \cite{ZhouJ.Barber1997} and the preparation of BZN films was not yet reported. In the BZT case, the authors observed that the films had low dielectric constant and high dielectric loss in comparison with the bulk ones, explaining the dependence of confocal Raman spectra, dielectric constant and dielectric losses on the annealing conditions with basis on the degree of thin film densification and Zn/Ta order in the crystalline grains. In this work we describe the synthesis of BZN thin films obtained by polymeric precursor method deposited by spin coating.

\section{Experimental details}
To prepare BZN thin films we used the polymeric precursors method using barium nitrate (Ba(NO$_3$)$_2$, Sigma Aldrich), zinc nitrate (Zn(NO$_3$)$_2$, Vetec), and ammonium complex of niobium (NH$_4$(NbO(C$_2$O$_4$)$_2$(H$_2$O)$_2$).3H$_2$O, CBMM) as metal sources. The barium polymeric precursor was obtained by dissolving 10 g of barium salt in 50 ml of distilled water. It was used the ratio 1:3 as molar ratio in the metal-citric acid. Citric acid (C$_6$H$_8$O$_7$.H$_2$O, Proqu\'{i}mico) was dissolved in distilled water and added to the solution of the metal salt kept stirring and heating to 60-70 \oC followed by ethylene glycol (HOCH$_2$CH$_2$OH, Merck) addition in the ratio 1:1 in relation to citric acid. The same procedure was employed to obtain the zinc polymeric precursor. In order to obtain the niobium polymeric precursor, it was dissolved 10 g of ammonium complex of niobium in 50 ml of distilled water under stirring and heating. Then led to the precipitation of niobium oxi-hydroxide until pH of 9 in a thermal bath at 0 \oC. Filtered vacuum to hold the niobium hydroxide (Nb(OH)$_5$) and elimination of oxalate ions with distilled water at 40-50 \oC. It is important to control the pH of the mixture to the same value. This step avoided precipitations of the precursor. We used the gravimetric analysis using a muffle furnace at 900 \oC for 1 h to determine the precipitate weight, in this case metal oxides obtained per gram of resin. With the gravimetric analysis, we determined the amount of each precursor polymer to obtain the mixture of the precursor BZN perovskite. After mixing, the three precursors on heating to 80-90 \oC formed a polyester precursor with high viscosity and glassy. The viscosity of the solution was adjust by water evaporation to obtain an acceptable range of viscosity values, 12$-$14 mPa$\cdot$s, measured at room temperature. BZN films were deposited onto Pt/Ti/SiO2/Si(100) substrates by spinning the deposition solution at 4.000 rpm for 20 s (spin coating technique). All the films were deposited layer by layer and a densification stage at 400\oC for 6h was employed after each layer deposition. A total of nine films were prepared. In six of them, we fixed the total number of layers (6 layers) and ranging the calcination time (2h, 4h, 8h, 16h, 32h, 64h). In the three remaining, we fixed the calcination time (6h) and ranging the number of layers (3, 6, 9 layers).

The crystalline phase of the films was investigated by X-ray diffraction (XRD$-$Bruker D8 Advance), in a continuous scanning mode using Cu-K$\alpha$ radiation, over a 2$\theta$ range 15$^\circ$-95$^\circ$. The XRD patterns were compared with data from ICSD (Inorganic Crystal Structure Database, FIZ Karlsruhe and NIST) International diffraction database (ICSD\#157044).

The confocal Raman spectra and the confocal images were acquired with an alpha 300 system microscope (Witec, Ulm, Germany), equipped with a highly linear (0.02\%) stage, piezo-driven, and an objetive lens from Nikon (100x, NA = 0.9). A Nd : YAG polarized laser ($\lambda$ = 532 nm) was focused with a diffraction-limited spot size (0.61$\lambda$/NA) and the Raman light was detected by a high sensitivity, back illuminated spectroscopic CCD behind a 1800 g/mm grating. The spectrometer used was a ultra-high throughput Witec UHTS 300 with up to 70\% throughput, designed specifically for Raman microscopy. The surface Raman image ($xy$ plane) was carried out in a region of 20 x 20 $\mu$m, with 60 points/line and 60 lines/image. The integration time in each point was 0.5 s. The region of the depth Raman image ($xz$ plane) was 10 $\mu$m wide and 20 $\mu$m deep, with 30 points/line and 30 lines/image. The integration time in each point was also 0.5 s.

The atomic force microscopy experiment was performed with an alpha 300 system microscope (Witec, Ulm, Germany), equipped with a contact-mode Al-coated cantilever with force constant of 0.2 N/m. The treatment of the AFM images was done using the WSxM software \cite{Horcas2007}.

\section{Results and discussions}

\begin{figure}
  \centering
  \includegraphics[width=\columnwidth]{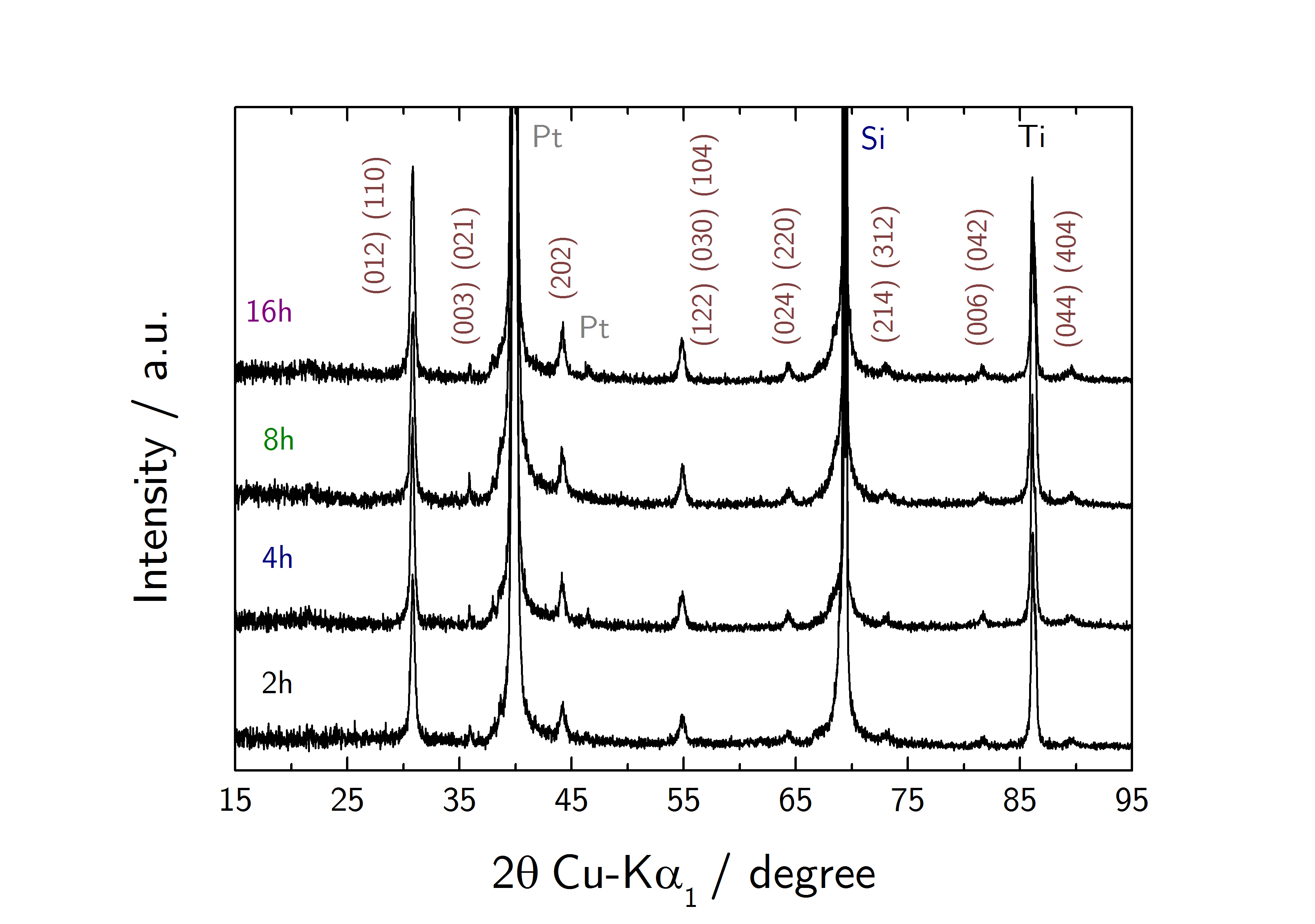}\\
  \caption{X ray diffractograms obtained for BZN thin films annealed at 900 \oC. The peaks associated to Silion, Platinum and Titanium are also indicates.}\label{BZNFilmDRX}
\end{figure}

Figure \ref{BZNFilmDRX} shows the X-ray diffraction patterns obtained for the all deposited films. The plane indexation was performed following the ordered trigonal \tri structure. It is important to point out that the peak around $2\theta=18^o$, which indicates the superstructure reflecting the ordering of Zn/Nb ions at B$'$/B$''$ sites,  was not clearly observed. This is due to the low diffraction intensity usually observed in thin films. Also, some partial disorder is expected due the high ZnO evaporation rates which difficult the obtainment of Zn compounds as showed by Varma et al by chemical methods \cite{Varma2006a}.

\begin{figure}
  \centering
  \includegraphics[width=\columnwidth]{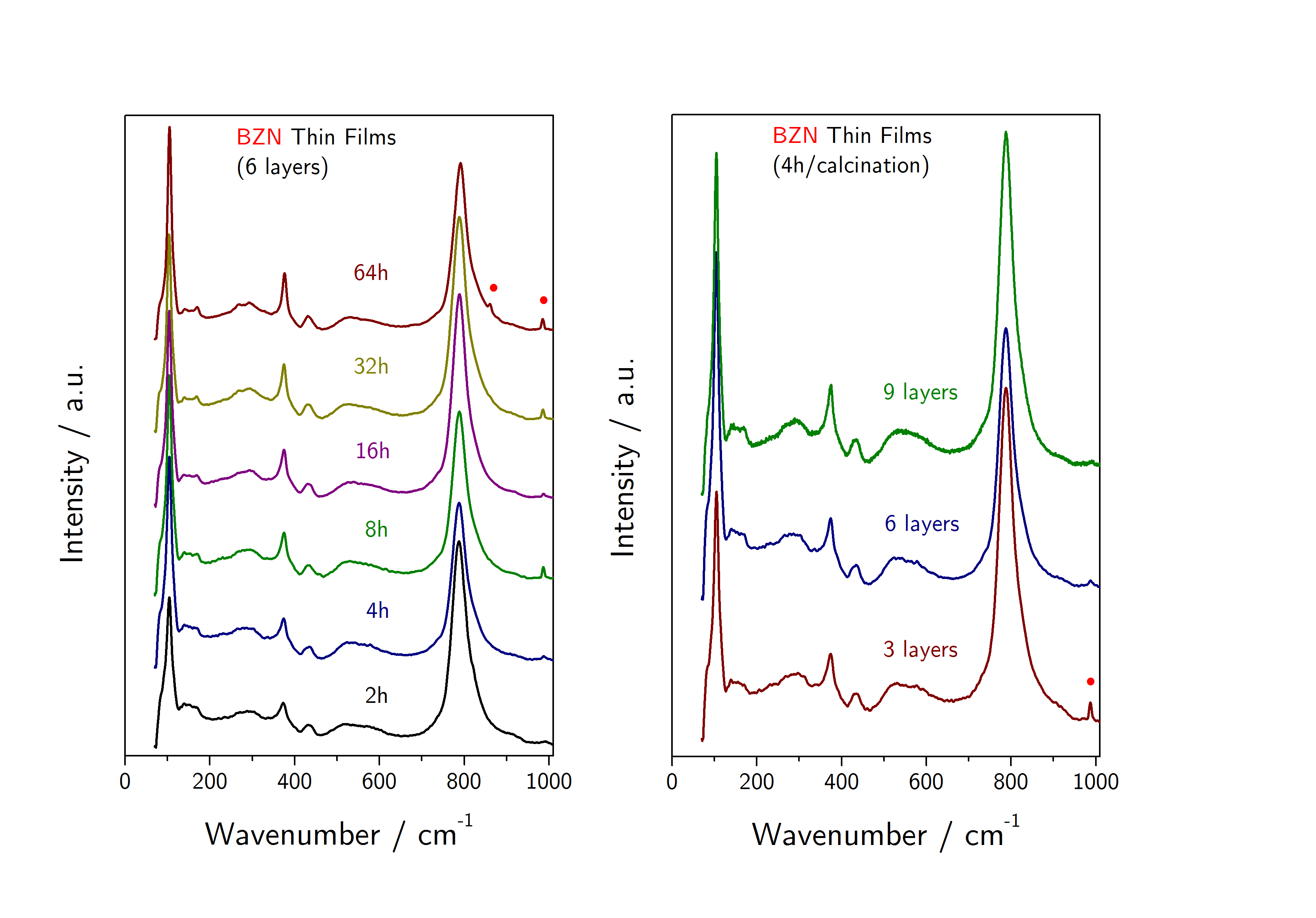}\\
  \caption{Raman spectra obtained for BZN thin films annealed at 900 \oC for several number and time of depositions.}\label{BZNFilmRaman}
\end{figure}

In order to observe the order and structure of the \BZN thin films we performed Confocal Raman measurements. Raman spectra obtained for \BZN thin films annealed at 900 \oC for several number and time of depositions are shown in Figure \ref{BZNFilmRaman}. The observed bands are shown in Table \ref{BZNFilmMODES} together with their assignments which were performed with basis in previous works \cite{Dias2003b,Dias2003c,Sagala1993a,SinyI.G.TaoR.KatiyarR.S.GuoR.Bhalla1998,Moreira2005,Dias2001,Moreira2001,MoreiraR.L.AndreetaM.R.B.HernandesA.C.Dias2005,Dias2007,Wang2009,Ning2012}. The spectra confirm the partially ordered trigonal structure assumed by the films. The local disorder is indicated by the modes 10(11) and 12(13), which are associated to the Nb(Zn) ion in the Zn(Nb) site \cite{Moreira2001}. However, although the X-ray measurements have not detected, the spectra also show the presence of BaNb$_2$O$_6$ as secondary phase. The peaks associated to this phase, that are indicated in the Figure \ref{BZNFilmRaman}, were observed at 860 \cm and 985 \cm. The observation of secondary phases in films deposited by chemical methods is usual \cite{DeJesus2010}. However, since X-ray technique has some limitations to detect crystalline phases with concentrations below 1\%, we estimated the concentration of BaNb$_2$O$_6$ lower than 1\%.
\begin{table}
  \centering
\begin{tabular}{cccc}
\hline
  Mode & Wavenumber / \cm & FWHM / \cm & Symmetry \\ \hline
1	  & 105	& 12  & $A_{1g}$      \\
2	  & 115	& 10  & $E_g$        \\
3	  & 144	& 21  & DAM   \\
4	  & 169	& 31  & $E_g$      \\
5	  & 266	& 37  & $A_{1g} $       \\
6	  & 295	& 32  & $E_g$      \\
7	  & 334	& 86  & FBL   \\
8	  & 376	& 14  & $E_g$        \\
9	  & 733	& 18  & $A_{1g}$      \\
10	  & 526	& 58  & $E_g$      \\
11	  & 578	& 81  & $E_g$   \\
12	  & 790	& 36  & $A_{1g}$   \\
13	  & 826	& 74  & $A_{1g}$   \\
14	  & 861	& 7  & IMP   \\
15	  & 985	& 6  & IMP   \\ \hline
\end{tabular}
  \caption{Observed Raman modes of BZN thin films and their respective symmetries. FBL, DAM and IMP indicate floating baseline, defect activated modes and modes due to the impurity, respectively.}\label{BZNFilmMODES}
\end{table}

\begin{figure}
  \centering
  \includegraphics[width=\columnwidth]{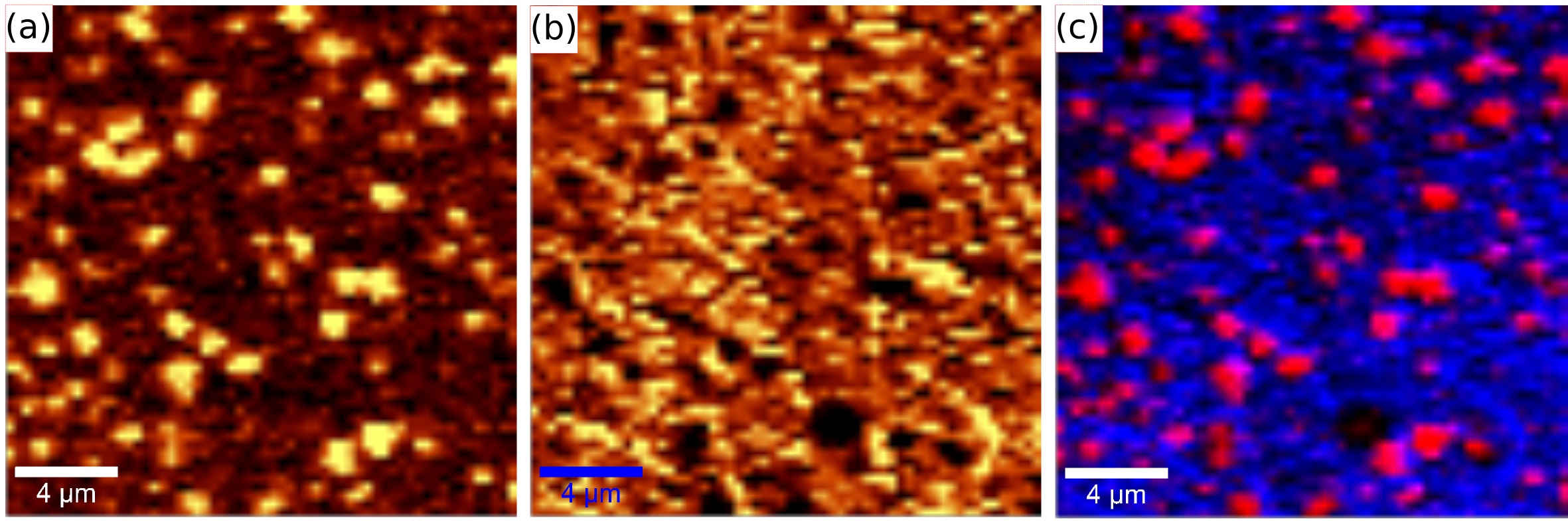}\\
  \caption{Spatial Raman maps of the modes at (a) 860 and (b) 100 \cm{} of \BZN{} thin films annealed at 900 $^\circ$C for 64 h. (c) Superposition of the two images on the left.}\label{BZNFilmMap}
\end{figure}

In order to determine the distribution of this secondary phase in the film, we have performed a Raman mapping on the film surface ($xy$ plane), as shown in Figure \ref{BZNFilmMap}. The spatial distributions of the BZN film and the secondary phase are shown in Figure \ref{BZNFilmMap}a and Figure \ref{BZNFilmMap}b, considering the peaks at 860 \cm{} and 100 \cm, respectively. So, as we can observe, there is a homogeneous distribution of the film, acting like a substrate (Figure \ref{BZNFilmMap}b) where some spots of the spatially localized secondary phase can be found. Although it is not shown, it is worth commenting that the Raman image of the peak at 985 \cm{} is very similar to that of Figure \ref{BZNFilmMap}a, confirming that the peaks at 860 \cm{} and 985 \cm{} have the same origin (second phase, BaNb$_2$O$_6$). Comparing images (a) and (b) of Figure \ref{BZNFilmMap}, it can be noted that they are complementary, indicating that the inclusion of one phase into another may be only physical. Figure \ref{BZNFilmMap}c summarizes the two images on the left and clearly shows their complementary pattern and the distribution of BaNb$_2$O$_6$ on the BZN film in a region of 20 x 20 $\mu$m.

\begin{figure}
  \centering
  \includegraphics[width=\columnwidth]{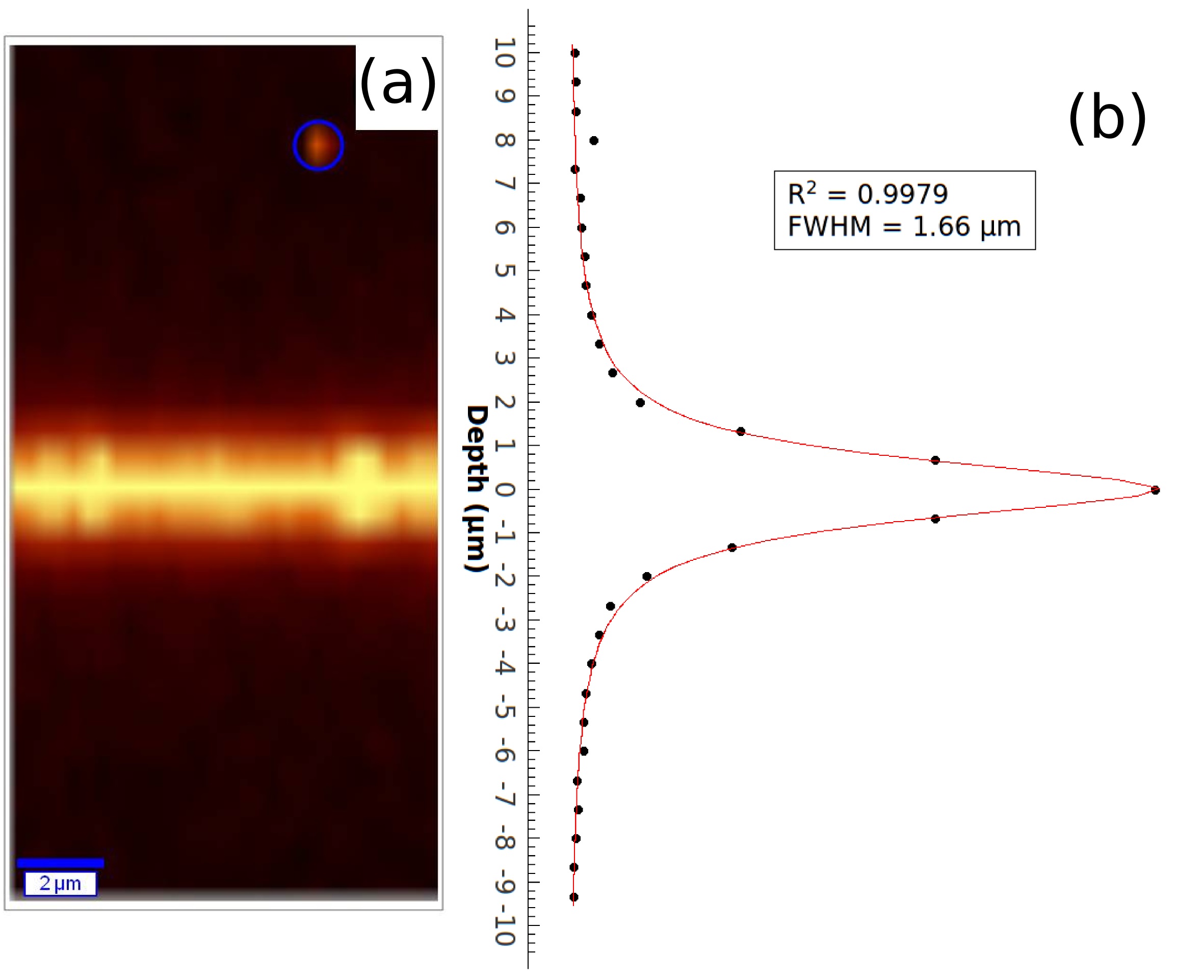}\\
  \caption{(a) Depth ($xz$ plane) confocal Raman image of the peak around 100 \cm{} obtained for \BZN{} thin films annealed at 900 $^\circ$C for 4 h with six deposition layers. (b) Sum of all the vertical profiles of (a), fitted with a Lorentzian curve.}\label{BZNFilmPROFILE}
\end{figure}

\begin{figure}
  \centering
  \includegraphics[width=\columnwidth]{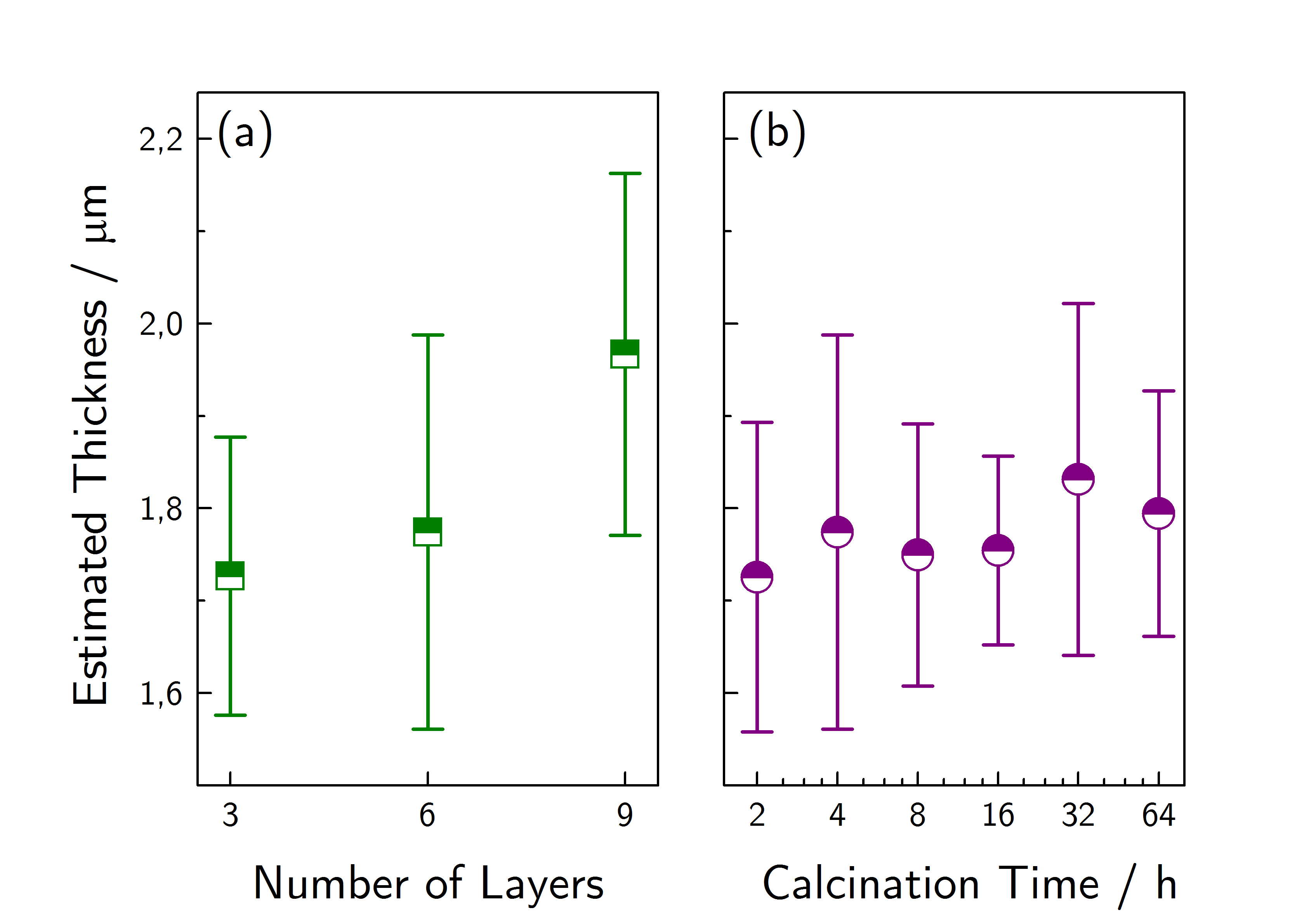}\\
  \caption{Estimated thin films tickness obtained for \BZN thin films annealed at 900 \oC in function of the (a) Number of depostions and (b) time of calcination.}\label{BZNFilmTHICKNESS}
\end{figure}

Aiming to estimate the film thickness, a depth confocal Raman measurement ($xz$ plane) was performed so as to map of the area under the vibrational mode with symmetry $A_{1g}$ observed around 100 \cm. This depth measurement was carried out first adjusting the focus of the objetive lens on the sample surface and then starting the experiment 10 $\mu$m above. The total depth was 20 $\mu$m with 30 points per line and 30 lines per image. This brief introduction of the experiment explains why the maximum of intensity of the peak at 100 \cm{} is found around $z=0$ $\mu$m (Figure \ref{BZNFilmPROFILE}a). The red spot inside a blue circle in Figure \ref{BZNFilmPROFILE}a is an experimental artifact and should be ignored. Summing all the 30 vertical lines of Figure \ref{BZNFilmPROFILE}a will introduce some statistics in the analysis of the data. The final depth profile, fitted with a Lorentzian curve, is shown in Figure \ref{BZNFilmPROFILE}b. This procedure estimated the film thickness as about 1.7 $\mu$m. The estimated thickness for all films are shown in Figure \ref{BZNFilmTHICKNESS}. We can observe that the film thickness, as expected, increases when the deposition number increases. However, the annealing time do not change significantly the thickness. Finally, it is important to comment that the films are transparent, minimizing the influence of the opacity in the experiment.

\begin{figure}
  \centering
  \includegraphics[width=\columnwidth]{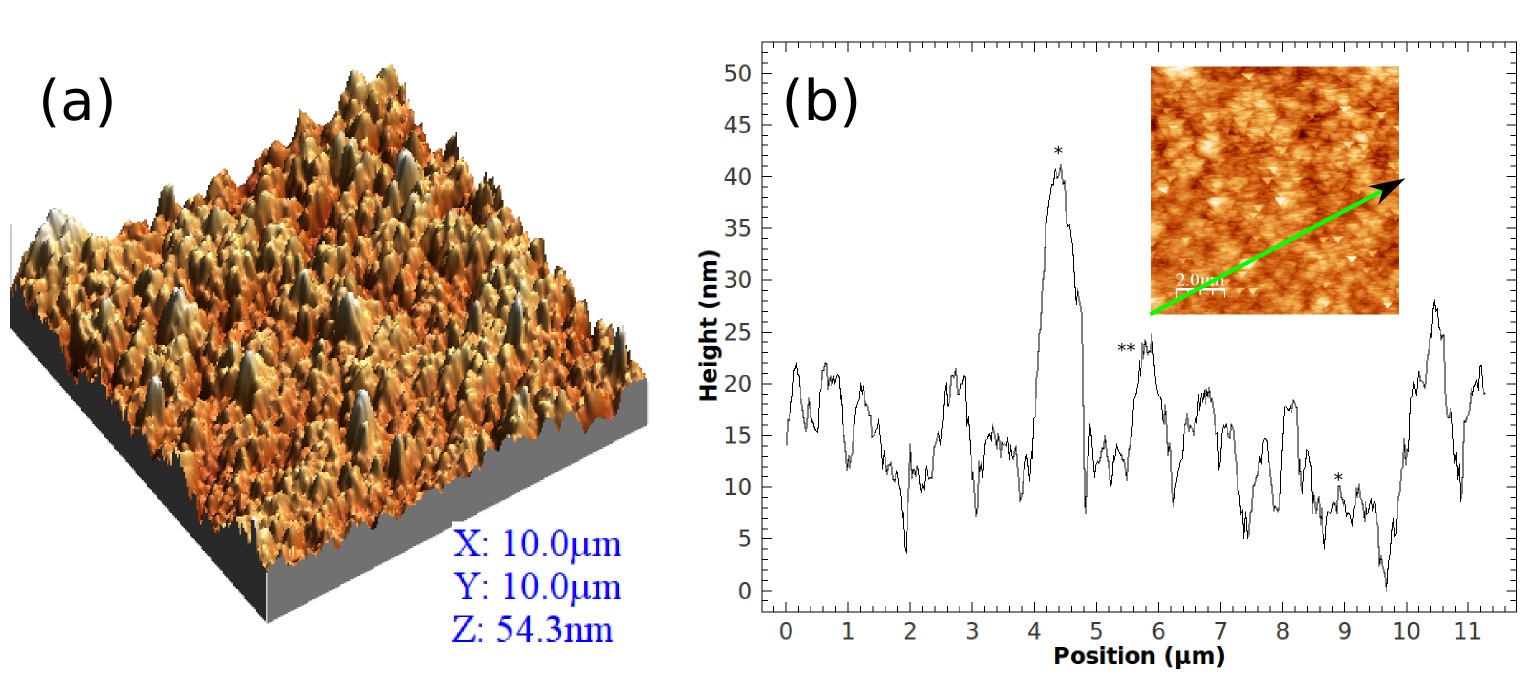}\\
  \caption{(a) Typical AFM micrography obtained for \BZN{} thin films annealed at 900 $^\circ$C for 4 h with six deposition layers. (b) Surface profile along the green arrow indicated in the inset. The inset is a front view of (a).}\label{BZNFilmAFM}
\end{figure}

The morphology of the film surface was investigated by AFM (Figure \ref{BZNFilmAFM}). It can be observed that the solvent evaporation induces a rich formation of peaks and valleys (Figure \ref{BZNFilmAFM}a). A variation of the height and the profile of the peaks along a path is shown in Figure \ref{BZNFilmAFM}b, where the inset shows a front view of Figure \ref{BZNFilmAFM}a and the path considered is indicated. It can be observed in Figure \ref{BZNFilmAFM}b that the heights of the peaks range from 10 to 40 nm (relative sizes 3 and 35 nm, respectively), as indicated by the one-star peaks. However, if the whole area is considered, the average value of the absolute height is 24.8 nm (two-stars peak). The roughness may be estimated by the average roughness ($S_a$) and the root mean square roughness ($S_q$), whose values are found to be 4.88 and 6.22 nm, respectively. An idea about the distribution of the sample height data can be given by the skewness ($R_{sk}$) and kurtosis ($R_{ku}$) parameters: 0.32 and 3.50. The former value indicates that the sample surface is approximately symmetric while the later indicates that our sample surface's peaks remind a normal distribution ($R_{ku}=3$). The estimated rugosity parameters obtained for the film morphology are summarized in Table \ref{BZNFilmRUGOSITY}.

\begin{table}
  \centering
  \begin{tabular}{cc}
    \hline
    Parameter & Values \\ \hline
    $S_a$ & 4.88 nm  \\
    $S_q$ & 6.22 nm \\
    $R_{sk}$ & 0.32 \\
    $R_{ku}$ & 3.50 \\
    \hline
  \end{tabular}
  \caption{Estimated rugosity parameters obtained for \BZN thin films annealed at 900 $^\circ$C for 4 h with six deposition layers.}\label{BZNFilmRUGOSITY}
\end{table}

\section{Conclusions}
In this works we investigated the \BZN thin films obtainment by polymeric precursors method. The results indicates that \BZN thin films show a partially disordered trigonal structure which was probed by X-ray diffraction and Raman spectroscopy measurements. Raman spectroscopy was abled to observe BaNb$_2$O$_6$ as secondary phase whose concentration was estimated lower than 1\%. Confocal Raman mapped the secondary phase and estimated the film thickness at around 2 $\mu$m. The film thickness increased with the deposition number and remain practically constant for all calcination time. The morphology of the film surface was characterized by atomic force microscopy which indicated a rough surface due to the solvent evaporation.

\section*{acknowledgement}
The authors are grateful to the Brazilian funding agencies CAPES, CNPq, and FAPEMA.


%

\end{document}